# Magnetic Fields in the Universe

Kandaswamy Subramanian
Inter-University Centre for Astronomy and Astrophysics
Post Bag 4, Ganeshkhind, Pune 411007

## Abstract

*The origin and maintenance of large scale magnetic fields in the Universe is discussed. A popular idea is that cosmic batteries generate seed magnetic fields which were amplified by magnetic induction due to turbulent motions, at times combined with differential rotation. We outline a seed field mechanism, the Biermann battery and then consider some basic ideas behind both small and large-scale turbulent dynamos. The small-scale dynamo may help understand magnetism in galaxy clusters and young galaxies, while the large-scale dynamo is important for the generation of fields with scales larger than stirring, as observed in nearby disk galaxies. Another intriguing possibility is that magnetic fields originated at some level from the early universe.*

## Introduction

The universe is magnetized. We all know that the Earth hosts a dipole magnetic field of order a Gauss which has been sustained for billions of years[1] and allows us to navigate using the compass. Our Sun is also magnetized, with its poles reversing very regularly, every 11 years[2,3]. The magnetic field of the Earth and the Sun are discussed elsewhere in this volume. On larger scales, magnetic fields of order a few micro Gauss and ordered on scales of a few to ten kilo parsecs (kpc) are detected in disk galaxies[4]. Recall that a parsec is about 3 light years and thus although the strength of this field is about a million times smaller than the Earth's magnetic field, it fills a very large volume and carries a substantial energy. Our Milky Way galaxy is itself such a disk galaxy. These galaxies contain about $10^{11}$ stars, which along with their interstellar medium, are in a thin disk supported against gravity by their rotation. Much less is known about magnetic fields in the other major type of galaxies, the ellipticals. Magnetic fields of similar strengths and coherence are detected even in the hot plasma filling the most massive collapsed objects in the universe, rich clusters of galaxies[5] and in young galaxies[6] which are several billion years younger than the Milky Way. Most recently there is indirect evidence that very weak magnetic fields a billion times smaller than that detected in galaxies could be filling the plasma in the void regions between galaxies[7]. The origin and maintenance of such cosmic magnetism is a fascinating issue of fundamental importance to modern astrophysics.

## The Induction equation

Magnetic fields decay either due to resistance which dissipates currents or because the Lorentz force exerted by the field drives motions and transfers magnetic energy to kinetic energy, which is in turn dissipated by viscosity. However, motions of a conducting plasma in a magnetic field can induce an electric field and if this field has a curl, from Faraday's law, the magnetic field can itself be maintained. The resulting evolution of the magnetic field **B** is given by the induction equation

$$\frac{\partial \boldsymbol{B}}{\partial t} = \nabla \times (\boldsymbol{V} \times \boldsymbol{B}) + \eta \nabla^2 \boldsymbol{B}$$

where **V** is the fluid velocity and $\eta$ the resistivity of the plasma. The induction equation implies that if **V** is zero, the field decays due the resistive term while if $\eta$ tends to zero the magnetic flux through any surface moving with the fluid is preserved independent of time, a statement known as Alfven's theorem. The dimensionless magnetic Reynolds number $R_m = V L/\eta$, where V and L are typical values of the fluid velocity and length scale in the system, measures the relative importance of induction (first term) and resistivity (second term). For most astrophysical systems $R_m$ is very large and the magnetic field is mostly frozen in to the plasma. We note that induction can only maintain the magnetic field only if there already exists a field, and thus we need a cosmic battery effect to generate a seed magnetic field which can be amplified. Battery effects usually generate a field much smaller than that observed. Therefore, we also need a velocity field **V** which can amplify and maintain the magnetic field, a process known as a dynamo. We shall touch upon both these processes below.

**The origin of seed magnetic fields**

How does one produce a current and a seed magnetic field in a plasma which originally had none? The Universe is charge neutral. However positively and negatively charged particles do not have identical properties. For example, if one considered a gas of ionized hydrogen plasma, then the electrons have a much smaller mass compared to protons. Thus for a given force acting on the electron and proton components of the plasma, the electrons will be accelerated much more than the protons. Such a force could be due to gradients in the pressure in the plasma. Note that pressure depends on density and temperature and if these are the same for electrons and protons, the pressure gradient force on these components will also be identical. The relative acceleration of electrons with respect to protons leads to an electric field, which couples back positive and negative charges so that they move together. Such a thermally generated electric field **E** is given by balancing the forces on the electrons, due to pressure gradient and the electric field assuming the protons are much more massive and so do not move. This gives $E = -\nabla p_e/en_e$, where $n_e$ is the electron number density and $p_e$ its pressure. We recall from Faraday's law of induction, that the curl of the electric field is proportional to the rate of change of the magnetic field. Thus if the above thermally generated **E** has a curl, then magnetic fields can grow from zero. The curl of a gradient is zero, and so the above electric field will have a curl only if the pressure (or temperature) gradient is nonaligned with the density gradient. Such a situation can indeed obtain in a rotating star with circulating velocity fields in their interior, and Biermann[8] first proposed such a battery effect to generate stellar magnetic fields.

The Biermann battery can also act in the cosmological context, where such non-parallel density and temperature gradients can arise for example when the first ultraviolet photon sources, like star bursting galaxies and quasars, turn on to ionize the intergalactic medium (IGM)[9]. The temperature gradient is normal to the front of ionization sweeping across the IGM, however the density can have gradient in a different direction, determined by arbitrarily laid down density fluctuations, which will later collapse to form galaxies and clusters. Such density fluctuations, need not be correlated to the source of the ionizing photons. The resulting thermally generated electric field then has a curl and magnetic fields correlated on galactic scales can grow. After compression during galaxy formation this process leads to a very small seed magnetic field[9] of order $10^{-21}$ Gauss, which though tiny, is a field which the dynamo can work on! This scenario has in fact been confirmed in detailed numerical simulations of the reionization of the IGM[10]. The Biermann battery can also operate in cosmological shocks which arise during the formation of galaxies and large scale structures to generate both vorticity and magnetic fields[11].

**Turbulent dynamos**

How does the seed magnetic field get amplified? One such possibility is turbulence or random motions, which is ubiquitous in all systems from stars to galaxy clusters. Turbulence arises due to instabilities like convection in stars, or is driven by other forces, like supernovae in galaxies, and sub-cluster mergers in the context of the intra cluster medium. The presence of turbulence in a highly conducting fluid generically leads to dynamo action, a process which is called the `turbulent dynamo'. Its analysis uses statistical methods or direct numerical simulations. Such dynamos are conveniently divided into two classes, the small- and large-scale dynamos, depending respectively on whether the generated field is ordered on scales smaller or larger than the scale of the turbulent motions.

**Small-scale turbulent dynamos**

The small-scale dynamo generates magnetic fields ordered on the energy carrying scales of the turbulent motions or even smaller. As we discussed above, the magnetic field is frozen into the fluid in a highly conducting plasma. And in any turbulent flow, fluid parcels random walk away from each other and so magnetic field lines get extended. Such random stretching is associated with compressions in the perpendicular directions, to conserve volume in nearly incompressible flows. Magnetic flux conservation then implies more field lines go through a given area, amplifying the field. This would continue until the field is concentrated into narrow enough regions that resistivity comes into play. At this stage the rate of field growth due to random stretching becomes comparable to resistive dissipation. What happens after this can only be addressed by a quantitative calculation. For an idealized random flow with very short correlation time, Kazantsev[12] showed that magnetic field can grow provided the Reynolds number $R_m$ associated with the turbulence is above a modest value of order 100. This is easily satisfied by the turbulent plasma in galaxies and galaxy clusters. The field then grows on the eddy turnover time $t_0 = l_0/v_0$. In galaxy clusters for a $v_0$ of order 100 km/s and $l_0$ of 10 kpc, we have $t_0 = 10^8$ yr, which is much smaller than the ages of galaxy clusters, which are a few billion years old at present. Similarly, for galactic plasma $v_0 = 10$ km/s and $l_0=100$ pc, $t_0 = 10^7$ yr again much smaller than the lifetime of even young galaxies. Thus the small-scale dynamo would amplify magnetic fields significantly in these systems. It perhaps provides the sole means for cluster magnetization and the first fields in young galaxies. The resulting field is however highly intermittent, especially when it is weak and active research is on to understand whether the degree of field coherence is sufficient to explain observations of the magnetic fields of clusters and young galaxies[13,14].

**Large-scale dynamos and galactic magnetism**

A remarkable change in the turbulent dynamo action occurs if the turbulence is helical or cyclonic, that is the velocity field has screw like motions with predominantly one sign for the handedness of the screw. In any rotating and stratified medium turbulence naturally becomes helical, like the generation of cyclones in the earth's atmosphere. This is because in such a medium, rising (falling) parcels of fluid expand (contract) and are twisted by the Coriolis force leading to helical motions with a net helicity. In the presence of helicity magnetic fields coherent on scales larger than the scale of the motions can develop. Suppose one starts with an initial large-scale field in the toroidal direction, that is, the $\phi$ direction circling the rotation axis of a star or a disk galaxy. Helical turbulent motions of the gas draw out the toroidal field into a loop and twists it to look like a twisted Omega. Such a twisted loop is connected to a current which has a component parallel to the original toroidal field. If the motions have a non-zero net helicity, this parallel component of the current adds up coherently. A toroidal current then results from the toroidal field. Hence, poloidal fields (in the r-z directions) can be generated from toroidal ones. This effect of helical turbulence is known as the $\alpha$ effect. Of course microscopic diffusion is essential to make permanent changes in the field. The possible importance of helical turbulence for large-scale field generation was first proposed by Parker[15].

The toroidal component of the field itself can be further enhanced by the shearing and stretching of the radial component of the poloidal field, by the differential rotation of the star or galaxy. This closes the toroidal-poloidal cycle and in disk galaxies leads to exponential growth of the large-scale (or mean) field, whose coherence scale can be much larger than the scale of the turbulence. Such a turbulent helical dynamo is thought to be the underlying mechanism for the generation and maintenance of large-scale magnetic fields in planets, stars and disk galaxies.

**Mathematics of mean-field dynamo theory**

A mathematical formulation of the mean-field (or large-scale) dynamo theory is now part of many text books and reviews[16,17,18]. Suppose the velocity field is split as $\mathbf{V} = \overline{\mathbf{V}} + \mathbf{v}$, into the sum of a mean large-scale velocity $\overline{\mathbf{V}}$ and a fluctuation velocity $\mathbf{v}$. The induction equation becomes a stochastic partial differential equation. The magnetic field is also split in a similar manner $\mathbf{B} = \overline{\mathbf{B}} + \mathbf{b}$ into a mean field $\overline{\mathbf{B}}$ and a fluctuation $\mathbf{b}$. Here the mean is a spatial average over scales larger than the turbulent eddy scales (but smaller than the system size) or an ensemble average and the mean of $\mathbf{v}$ and $\mathbf{b}$ is zero. Taking the average of the induction equation one gets the mean-field or large-scale dynamo equation

$$\frac{\partial \overline{\mathbf{B}}}{\partial t} = \nabla \times (\overline{\mathbf{V}} \times \overline{\mathbf{B}} + \boldsymbol{\epsilon}) + \eta \nabla^2 \overline{\mathbf{B}} \ .$$

This large-scale dynamo equation has now importantly a new term, the mean turbulent electromotive force $\boldsymbol{\epsilon} = \overline{\mathbf{v} \times \mathbf{b}}$ which is determined by the statistical correlation between the fluctuating $\mathbf{v}$ and $\mathbf{b}$. A central problem of mean-field dynamo theory is to express this emf in terms of the mean field itself. For isotropic, homogeneous, helical 'turbulence' in the approximation that the correlation time is short one finds the $\boldsymbol{\epsilon} = \alpha \overline{\mathbf{B}} + \eta_t \nabla \times \overline{\mathbf{B}}$ where $\alpha$ depends on the helicity of the random velocity and $\eta_t$ on its energy per unit mass $\overline{v^2}$. Note that the $\alpha$ effect is the mathematical representation of what we described above as the effect of helical motions. Also $\eta_t$ simply adds to the microscopic diffusion in the dynamo equation and is called turbulent magnetic diffusion. In disk galaxies $\overline{\mathbf{V}}$ is that of differential rotation of the disk where the rotation frequency $\omega$ decreases with radius. Its effect, sometimes referred to as the $\omega$ effect, is to shear radial fields to toroidal fields as described qualitatively above, while the $\alpha$ effect is crucial for regeneration of poloidal from toroidal fields. Their combined effect can lead to the grow th of the disk galaxy magnetic field and hence a dynamo provided they can overcome the destructive effect of turbulent diffusion. In disk galaxies one finds[19] that the large-scale field typically grows on time scales a few times the rotation time scale, of order $10^8$ - $10^9$ yrs, again much smaller than galactic ages of order $10^{10}$ yrs.

This picture of large-scale dynamo action faces several potential problems. Firstly, in a highly conducting fluid, magnetic helicity, which measures the linkages between field lines is nearly conserved. Then when helical motions are writhing say the toroidal field to generate a poloidal field, an oppositely signed twist is developing on smaller scale, to conserve magnetic helicity. The Lorentz force associated with this small-scale twist goes to oppose the writhing effect of helical motions and quench the dynamo. Large-scale dynamos can only then work by shedding this small-scale magnetic helicity[18,20,21]. Mean-field dynamo models incorporating magnetic helicity conservation have been used to understand a wide variety of observations of galactic magnetism[22,23].

Further, while the mean-field dynamo operates to generate the large-scale field, the small-scale dynamo is producing small-scale fields typically at a much faster rate. When the small scale field grows to be comparable with the turbulent kinetic energy, we expect Lorentz forces to come into play and saturate the field growth. Can then the mean field dynamo continue to operate to grow fields? Direct simulations of turbulent helical dynamos[24] find that initially scales both larger and much smaller than the stirring scale grow together dominated by small scales, but crucially on saturation due to the Lorentz force, larger and larger scales come to dominate, eventually leading to system scale fields, provided magnetic helicity can be efficiently removed. These issues remain at the forefront of current research on turbulent dynamo theory.

**Magnetic fields from the early universe**

Could magnetic fields be a relic from the early Universe, arising perhaps during the inflationary epoch when density fluctuations are also generated or in some other phase transition? Indeed, such a mechanism would be favored, if the tentative indications, that magnetic fields at the femto Gauss level pervade even the voids regions almost empty of galaxies, are firmed up. In the late universe void regions can only be magnetized if dynamos operate efficiently in say galaxies and then energetic magnetized winds are driven out from galaxies to fill the voids. On the other hand if the universe can be magnetized at early epochs, the field would pervade all the volume including regions which will become voids. Such processes are also of interest as they could provide a handle to probe the physics of the early universe.

Interestingly, primordial magnetic fields with a present day strength of a nano Gauss and coherent on galactic and larger scales can strongly influence a number of astrophysical processes[25,26]. For example, they lead to additional temperature and polarization anisotropies in the Cosmic Microwave Background (the relic radiation from the very early universe), induce dwarf galaxy formation at high redshifts and also partially reionize the universe. They can also provide a strong seed field to the dynamo. The energy density of the cosmic microwave background is equivalent to the energy in a magnetic field of about 3 micro Gauss. To generate fields at the nano Guass level at present, only a fraction of about $10^{-7}$ of the energy density in the early universe needs to be channelized into coherent magnetic fields and an even smaller fraction to explain the possible fields in voids.

However, this does not turn out to be trivial. During inflation, the Universe is undergoing rapid accelerated expansion, and quantum vacuum fluctuations of the electromagnetic (or more correctly the hyper magnetic) field can be naturally excited and transformed to classical fluctuations. Large-scale magnetic fields result, by the rapid expansion of the scale of these fluctuations. But the same expansion dramatically dilutes magnetic flux if standard electromagnetism is obeyed. One has to break what is known as the conformal (or scale) invariance of electromagnetism[27]. The strength of the generated field then depends sensitively on how exactly one does this. Electroweak or QCD phase transitions, which occur as the plasma in the early Universe cools, can lead to appreciable field strengths, where a significant fraction of the energy density of the Universe is channelized into magnetic fields. However, this being a causal process can only generate fields on the very small coherence scales limited by the Hubble radius. These fields decay and order themselves as the universe expands. For them to be sufficiently strong at present, an inverse cascade of magnetic energy to larger scales is required[28,29].

In summary, it appears that the origin of cosmic magnetism on the largest scales of galaxies, galaxy clusters and the general inter galactic medium is still an open problem. Indeed, the origin of cosmic magnetism is one of the key goals of upcoming radio telescopes like the Square Kilometer Array[30]. A recent added bonus is the realization that gamma-ray astronomy could also probe very weak intergalactic magnetic fields[7]. These features make cosmic magnetism an attractive and challenging field for young minds to work on!

1. P. Olson, The geodynamo's unique longevity, Physics Today, **66**, 30 (2013)
2. D. H. Hathaway, The Solar Cycle, Living Reviews in Solar Physics, **7**, 1 (2010)
3. A.R. Choudhuri, Nature's Third Cycle: A Story of Sunspots, Oxford University Press (2015)
4. R. Beck, Magnetic fields in spiral galaxies, The Astronomy and Astrophysics Review, **24**, 4 (2015)
5. F. Govoni, L. Feretti, Magnetic fields in clusters of Galaxies, International Journal of Modern Physics D, **13**, 1549 (2004)
6. M. L. Bernet, F. Miniati, S. J. Lilly et. al., Strong magnetic fields in normal galaxies at high redshift, Nature, **454**, 302 (2008)
7. A. Neronov, I. Vovk, Evidence for Strong Extragalactic Magnetic Fields from Fermi Observations of TeV Blazars, Science, **328**, 73 (2010)
8. L. Biermann, Uber den Ursprung der Magnetfelder auf Sternen und im interstellaren Raum, Zeitschrift Naturforschung, **A5**, 65 (1950)
9. K. Subramanian, D. Narasimha, S. M. Chitre, Thermal generation of cosmological seed magnetic fields in ionization fronts, Monthly Notices of Royal Astronomical Society, **271**, L15 (1994)
10. N. Y. Gnedin, A. Ferrara, E. G. Zweibel, Generation of the Primordial Magnetic Fields during Cosmological Reionization, The Astrophysical Journal, 539, 505 (2000)
11. R. M. Kulsrud, R. Cen, J. P. Ostriker, D. Ryu, The Protogalactic Origin for Cosmic Magnetic Fields, The Astrophysical Journal, **480**, 481 (1997)
12. A. P. Kazantsev, Enhancement of a Magnetic Field by a Conducting Fluid, Soviet Physics JETP, 26, 1031 (1968)

13. P. Bhat, K. Subramanian, Fluctuation dynamos and their Faraday rotation signatures, Monthly Notices of the Royal Astronomical Society, **429**, 2469 (2013)
14. S. Sur, P. Bhat, K. Subramanian, Faraday rotation signatures of fluctuation dynamos in young galaxies, Monthly Notices of the Royal Astronomical Society, **475**, L72 (2018)
15. E. N. Parker, Hydromagnetic dynamo models, The Astrophysical Journal, **122**, 293 (1955)
16. H. K. Moffatt, Magnetic field generation in electrically conducting fluids, Cambridge University Press (1978)
17. F. Krause, K. H. Radler, Mean-field magnetohydrodynamics and dynamo thory, Academic-Verlag, Berlin; also Pergamon Press, Oxford (1980)
18. A. Brandenburg, K. Subramanian, Astrophysical magnetic fields and nonlinear dynamo theory, Physics Reports, **417**, 1 (2005)
19. A. Shukurov, Introduction to galactic dynamos, in `Mathematical aspects of natural dynamos', eds. E. Dormy and B. Desjardins, EDP Press (arXiv:astro-ph/0411739) (2004)
20. E. G. Blackman, Magnetic Helicity and Large Scale Magnetic Fields: A Primer, Space Science Reviews, **88**, 59 (2015)
21. K. Subramanian, A. Brandenburg, Magnetic helicity and its flux in weakly inhomogeneous turbulence, Astrophysical Journal Letters, 648, L71 (2006)
22. L. Chamandy, K. Subramanian, A. Shukurov, Galactic spiral patterns and dynamo action - I. A new twist on magnetic arms, Monthly Notices of the Royal Astronomical Society, **428**, 3569 (2013)
23. L. Chamandy, A. Shukurov, R. A. Taylor, Statistical Tests of Galactic Dynamo Theory, The Astrophysical Journal, **833**, 43 (2016)
24. P. Bhat, K. Subramanian, A. Brandenburg, A unified large/small-scale dynamo in helical turbulence, Monthly Notices of the Royal Astronomical Society, **461**, 240 (2017)
25. R. Durrer, A. Neronov, Cosmological magnetic fields: their generation, evolution and observation, The Astronomy and Astrophysics Review, 21, 62 (2013)
26. K. Subramanian, The origin, evolution and signatures of primordial magnetic fields, Reports of Progress in Physics, **79**, 076901 (2016)
27. M. S. Turner, L. M. Widrow, Inflation-produced, large-scale magnetic fields, Physical Review D, **37**, 2743 (1988)
28. R. Banerjee, K. Jedamzik, Evolution of cosmic magnetic fields: From the very early Universe, to recombination, to the present, Physical Review D, **70**, 123003 (2004)
29. A. Brandenburg, T. Kahniashvili, A. G. Tevzadze, Nonhelical Inverse Transfer of a Decaying Turbulent Magnetic Field, Physical Review Letters, **114**, 075001 (2015)
30. B. M. Gaensler, R. Beck, L. Feretti, The origin and evolution of cosmic magnetism, New Astronomy Reviews, **48**, 1003 (2004)